# Preterm neonates distinguish rhythm violation through a hierarchy of cortical processing

Edalati, M.[1,2], Mahmoudzadeh, M.[1,3], Kongolo, G.[1,3], Ghostine, G.[1,3], Safaie, J.[2], Wallois, F.[1,3*], and Moghimi, S.[1,2,3*]

[1] Inserm UMR1105, Groupe de Recherches sur l'Analyse Multimodale de la Fonction Cérébrale, CURS, Avenue Laennec, 80036 Amiens Cedex

[2] Electrical Engineering Department, Ferdowsi University of Mashhad, 9177948974 Mashhad, Iran.

[3] Inserm UMR1105, EFSN Pédiatriques, CHU Amiens sud, Avenue Laennec, 80054 Amiens Cedex

\* Corresponding authors: Fabrice Wallois, Sahar Moghimi

fabrice.wallois@u-picardie.fr

sahar.moghimi@u-picardie.fr

# Preterm neonates distinguish rhythm violation through a hierarchy of cortical processing

**Abstract-** Rhythm is a fundamental component of the auditory world, present even during the prenatal life. While there is evidence that some auditory capacities are already present before birth, whether and how the premature neural networks process auditory rhythm is yet not known. We investigated the neural response of premature neonates at 30-34 weeks gestational age to violations from rhythmic regularities in an auditory sequence using high-resolution electroencephalography and event-related potentials. Unpredicted rhythm violations elicited a fronto-central mismatch response, indicating that the premature neonates detected the rhythmic regularities. Next, we examined the cortical effective connectivity underlying the elicited mismatch response using dynamic causal modeling. We examined the connectivity between cortical sources using a set of 16 generative models that embedded alternate hypotheses about the role of the frontal cortex as well as backward fronto-temporal connection. Our results demonstrated that the processing of rhythm violations was not limited to the primary auditory areas, and as in the case of adults, encompassed a hierarchy of temporo-frontal cortical structures. The result also emphasized the importance of top-down (backward) projections from the frontal cortex in explaining the mismatch response. Our findings demonstrate a sophisticated cortical structure underlying predictive rhythm processing at the onset of the thalamocortical and cortico-cortical circuits, two months before term.

**Introduction**

We are exposed to rhythmic patterns from the very early stages of prenatal development. As early as 25 weeks gestational age (wGA), the fetus is equipped with structural components of the auditory system that allow him to hear the isochronous rhythm of the maternal heartbeat and respiration (1, 2). He is also constantly exposed to environmental sounds originating from outside the uterus. Despite the attenuation of sound and distortion of its frequency by maternal tissue, the rhythmic information is preserved (3). Therefore, auditory rhythm processing might begin from the very early stages of development. However, the underlying neural circuits and mechanisms that process such information are undergoing rapid and dynamic evolution (4, 5). Neuronal migration, as well as synaptogenesis, occurs during the last trimester of gestation. The number of synapses in the subplate is high and its size continues to increase until 33 wGA. The first synapses appear in the cortical plate at approximately 26 wGA. At approximately 28 wGA, the thalamic afferents "waiting" in the subplate are relocated to layer IV of the cortex (5). This period is also marked by the rapid emergence of short-range connectivity, in addition to the long-range association pathways already observed (6). Despite their immature structure and function, these neural networks already create neural responses to the spectral characteristics of sound (7, 8).

Understanding the auditory world around us requires a fine perception of time. The temporal grouping of auditory events and perception of rhythmic patterns is central to how we recognize structure in sound sequences. Rhythm is a particularly important feature of acoustic events because its temporal regularity enables predictions about upcoming sounds. Such sequential predictive processes are essential from the developmental point of view and are fundamental abilities that need to be acquired during the course of early development to interpret both speech and music, as well as coordinate actions (walking and catching) (9). Evidence suggests that soon after birth, the newborn's brain is sensitive to the temporal pattern of sound (10, 11). Whether this capacity is present in newborns due to learning during the last months of pregnancy or evolution has bestowed upon us the genetically directed neural organization to process this aspect of the auditory world is still an open discussion. This study joins this debate by addressing the question of rhythm perception in premature neonates during their first week of life.

Predictive coding has furnished a unifying theory of the neural mechanisms underlying predictive processes, learning, and plasticity in auditory comprehension (12-14). In this framework, tone pattern learning and prediction encompass a process of optimizing an internal neural model of temporal regularities (15), which is engendered through reciprocal bottom-up and top-down neural circuits (16), and, at the cortical level, involves the primary auditory, auditory association, frontal, and motor cortices (17-19). When predictions fail, the brain generates "prediction error signals", which are processed through neural hierarchies to revise the model's prediction. One of the neural manifestations of prediction error in auditory processing is mismatch negativity (MMN), which has been repeatedly observed in response to the violation of rhythmic structures (20-30).

There is evidence for early sensitivity to the temporal organization of tone sequences in infants. Electroencephalography (EEG) and Magnetoencephalography in sleeping newborns show sensitivity to the onset, offset, and tempo of auditory sequences (10), the omission of metrically important tones in a rhythmic pattern (11), and hierarchical regularities (31). In addition, infants in the first year of life show neural entrainment to the beat and meter of rhythmic patterns (32) and detect violations of meter in an auditory sequence (33). Interestingly, preterm neonates produce a mismatch response to phoneme and voice deviations from a regular sequence (7, 8), suggesting that predictive processing is already present at this very early stage of neurodevelopment.

The perception of auditory temporal regularities over multiple events encompasses a reciprocal dialog between various cortical structures, including the primary auditory cortex, auditory association cortex, and frontal cortex (15, 34, 35), and is associated with higher order representation of the auditory environment in the frontal cortex (14). functional magnetic resonance imaging (fMRI) (36, 37) and functional near-infrared spectroscopy (fNIRS) (7) have demonstrated activity over the frontal regions during auditory tasks in the newborn and preterm brain. Thus, auditory perception and learning at this very early stage of neurodevelopment is not limited to low-level processes and may also occur along the predictive hierarchy involving higher level cortical regions (as for later stages of development (38)). Here, we first tested the hypothesis that the premature brain at between 30 to 34 wGA (two months before the equivalent age of term) detects the temporal regularities in a rhythmic pattern. Next, we tested whether this phenomenon already involves an effective network constituting various hierarchical cortical structures, including the frontal cortex, as well as the contribution of backward connections from the frontal cortex. The presence of backward connections has been linked to the generation of top-down predictions about future upcoming auditory events, a mechanism that has been demonstrated to be present in adults (16, 18). We tested this hypothesis quantitatively, first using high-resolution event-related potential (ERP) analysis in response to rhythmic deviations. Second, we applied dynamic causal modeling (DCM), which allows for inferences about the neuronal architecture underlying the generated electric signals and provides an efficient way to map from observed evoked potential patterns to causative neuronal mechanisms (39-42).

## Materials and Methods

### Participants

High-resolution EEGs were recorded for 20 (eight females) healthy preterm neonates with mean gestational age at birth, 31.48 ± 1.23 wGA (mean recording age: 33 ± 1.44 wGA) while sleeping, Table S.1. The EEGs were recorded in incubators at the neonatal intensive care unit of the Amiens University Hospital (Amiens, France). All neonates had appropriate birth weight, size, and head circumference for their term age, an APGAR score > 6 at 5 min, and normal auditory and clinical neurological assessments. None were considered to be at risk of brain damage. In particular, the results of a neurological examination at the time of the recordings had to correspond to the corrected gestational age, with no history of abnormal movements. The gestational age (estimated from the date of the mothers' last period and ultrasound measurements during pregnancy) was consistent with the degree of brain maturation (evaluated on the EEG). The brain imaging results (particularly transfontanellar ultrasound and standard EEG) had to be normal. One or both parents were informed about the study and provided their written informed consent. The local ethics committee (CPP Ouest I) approved the study (ID-RCB: 2019-A01534-53).

### Auditory stimuli and the experimental paradigm

The stimulus consisted of an auditory rhythm in 2/4 meter presented continuously at 60 beats per minute (1 s per quarter note). The inter-tone intervals were 1000 ms, 500 ms, and 500 ms for standard rhythm trials and 1000 ms, 250 ms, and 750 ms for deviant rhythm trials. We used a 990-Hz pure tone with a duration of 150 ms. All tones had a rise and fall time of 10 ms. The frequency content of the tones did not change, neither between the standard rhythm and deviant rhythm conditions nor between different trials. A dynamic accent of 25% above the general intensity was induced on the first beat of each stimulus (first tone of each trial) to reinforce the perceived meter. Stimuli were synthesized using the open-source software Audacity 2.2.2 (www.audacity.sourceforge.net) and exported as wav-files.

The stimuli were delivered in the context of an oddball paradigm (Fig.1A). The experimental session included three test sequences. In the main test sequence, the high-probability standard rhythm trials (p = 79%, 1,518 trials) were interspersed with the infrequent deviant rhythm trials (p = 21%, 400 trials). The order of presentation of the deviant rhythm trials was pseudorandomized among the standard rhythm trials, enforcing three to seven standard rhythm trials between successive deviant rhythm trials. Two additional control sequences were randomly interspersed in the main test sequence. These sequences had 150 deviant rhythm trials each (repeated 150 times without any standard trial in between). Stimuli were delivered through a speaker at 65 dB SPL, located at the neonates' feet (Fig.1B), using Psychtoolbox MATLAB (43). The total duration of the experiment was ~74 min.

### EEG acquisition and preprocessing

EEG signals were collected using a 124-channel HydroCel GSN net with an Electrical Geodesic NetAmps 200 amplifier passing a digitized signal to Electrical Geodesics NETSTATION software (v.5). Impedances were kept below 50 kΩ. The EEG was digitized at a 1,000-Hz sampling rate, with a Cz vertex electrode as reference. The recorded signals were analyzed with MATLAB® software (The MathWorks, Inc., Natick, Massachusetts, United States) using FieldTrip (44), EEGLAB (45), and custom MATLAB functions and codes. We applied a two-pass 0.5- to 45-Hz finite impulse response (FIR) filter (order = 3 cycles of the low-frequency cut-off) and a 50-Hz notch filter by EEGLAB toolbox to remove low- and high-frequency artifacts and also the line noise from the EEG signals. After down-sampling to 512-Hz, we removed 13-17 electrodes for each subject through visual inspection because of the low signal-to-noise ratio. Artifacts (e.g., ECG, eye movement, and muscle activity) were then removed by independent component analysis (ICA) using the EEGLAB toolbox. Trials were excluded if the standard deviation of amplitude exceeded 25 µV within two moving windows of 200 and 800 ms, the sample gradient exceeded 6 µV, the absolute amplitude remained below 0.1 µV, or any sampling point exceeded 60 µV at any electrode location. The aforementioned artifact rejection process was carried out individually for each electrode to remove contaminated trials for electrodes near the noise source and preserve the unaffected electrodes (46). We rejected the entire trial if 30% of the channels over the corresponding trial were rejected. In addition, we rejected a channel if 70% of the trials on the corresponding channel were rejected. Three subjects were discarded after this step due to the small number of remaining trials. The mean number of remaining trials for all neonates and all electrodes for the various conditions was $291.04 \pm 40.56$ (rhythm deviant), $216.89 \pm 29.51$ (rhythm control), and $993.82 \pm 344.27$ (standard). EEG data were later re-referenced to the average reference.

**Event-related potentials (ERPs)**

The data were low-pass FIR filtered at 16-Hz (13 cycles) and epoched, starting 100 ms before the onset of the deviant tone (third tone in the deviant trial) and ending 750 ms after, to investigate the ERP components. Detrending was applied to each epoch to remove the observed trend. Event-related potentials were computed by averaging the EEG trace of the detrended epochs corresponding to each condition after baseline ([-100 to 0] ms) correction. A nonparametric cluster-based permutation procedure (5,000 permutations), implemented in the FieldTrip toolbox (47), was applied to search for significant changes in the deviant condition relative to the control condition. The initial threshold for cluster definition was set to $p < 0.05$ and the minimum number of neighbors to 4. Finally, the final significance threshold for summed $t$ values within clusters was set to $p < 0.05$.

**Dynamic causal modeling (DCM)**

We evaluated our hypothesis about the hierarchical mechanisms underlying the mismatch response (MMR) in preterm neonates using DCM in SPM 12 (v.7771) to investigate the effective connectivity between temporal and frontal cortical sources, previously shown to participate in the generation of auditory MMN in studies carried out in adults (18, 48-51). The location of cortical sources, namely the bilateral primary auditory cortex (A1), superior temporal gyri (STG), and inferior frontal gyri (IFG), were specified by two experts using an MRI of a preterm neonate of 32 wGA (52). The forward projections of these sources to the sensors were modeled with specified parameters using our neonatal finite element head model (53), which consists of six compartments: white and gray matter, cerebrospinal fluid, fontanels, skull, and scalp. Data were detrended and reduced to eight spatial modes to reduce the computational load before model fitting (49). Sources of control and deviant trials were reconstructed separately using the forward modeling described above and inverted using the SPM 12 standard algorithm with default settings (49).

DCM makes inferences about the underlying mechanisms of ERP components and the corresponding changes in coupling between equivalent current dipole sources using biophysically constrained neural mass modeling (54, 55). It considers a group of specific models to provide evidence in favor of one model relative to others through Bayesian model selection (BMS) (56). We used 16 generative models (Fig. 2) to test our hypothesis concerning the mechanisms underlying the MMR in preterm neonates, more precisely, the effective connectivity between temporal and frontal sources for the time window of 0–375 ms from the onset of the deviant tone (encompassing the first MMR) and the possible involvement of the top-down connections in explaining MMR dynamics. All connections between MMR sources were bidirectional and modulated (42). The first four models were used to study the performance of models without the involvement of the IFG (NI family, models 1 to 4). The following six models were added to evaluate the hypothesis of the presence of only forward connections and therefore undeveloped backward connections between the STG and IFG in preterm neonates (forward (F) family, models 5 to 10). This hypothesis assumes the presence of bottom-up but absence of top-down transfer of information during the MMR. The last six models were replications of previous studies in adults with full forward and backward connections between sources, (forward-backward (FB) family, models 11 to 16) (18, 48, 49), involving the left and right IFG. This family of models assumed both bottom-up and top-down connections in the generation of the MMR. All models began with driving inputs into the bilateral primary auditory cortex, with or without intrinsic connections within these sources, followed by bidirectional connections to the bilateral STG and forward or forward-backward connections to unilateral or bilateral IFG.

We used BMS (56) to compare the generative models and discover which fit best with the observed neural responses. We employed two approaches to compare models: fixed-effects (FFX) and random-effects (RFX). Given the fact that, physiologically, one model should fit to all subjects, we first used FFX to determine the model that best explained the neural responses. FFX assumes that participants use the same network architecture but have varying connection strengths (18, 57). In addition, we applied the RFX approach to the models and families across subjects to take into consideration the possible bias of the FFX results due to participant outliers (18). The model/family with the highest log-evidence and exceedance probability was selected as the "winning" model. A strong model is usually a model with log-evidence at least three units above that of the other models (a three-unit difference corresponds to a Bayes factor of 20 and, by convention, is considered to be strong evidence for one model over another (57, 58)). The posterior probability corresponding to each model was also calculated to indicate the probability of the winning model given the neural responses within the current model space.

## Results

Rhythmic deviation induced neural MMRs in the cortical network of preterm neonates with a particular time course and topographical distribution. The analyses corresponding to DCM demonstrate that the MMR is best explained by a model with both forward and backward connectivity between temporal and frontal sources.

### Mismatch response to rhythm deviation

The ERP response to the deviant rhythm condition is depicted in Fig. 3. The time window [-100 to 0 ms] was considered to be the baseline. The grand average ERPs illustrate the MMR manifesting as enhanced early (~150-350 ms) frontal and fronto-central positivity for the deviant rhythm condition with respect to the control condition, consistent with the typical time window of the well-known MMN. A subsequent negative deflection proceeded the MMR in the 400- to 500-ms time window, followed by another late positive deflection that extended to the next trial. Both components were more pronounced over the frontal and fronto-central electrodes by visual inspection.

Cluster-based statistics showed two spatiotemporal clusters: a positive cluster ($p = 0.0028$, corrected), comprised of frontal and fronto-central electrodes extending approximately over 215-

301 ms and a negative cluster ($p = 0.01$, corrected), comprised of frontal electrodes extending approximately over 430-470 ms post-final tone. The approximate spatial and temporal extent of the two clusters are depicted in Fig. 3A. Fig. 3B shows the uncorrected t-values resulting from the comparison between the deviant and control conditions for each electrode and the epoch time window. The topographical distributions of the t-values for the significant time windows are shown in Fig. 3C.

**Effective cortical connectivity underlying the deviant rhythm response using DCM**

We used DCM modeling in an attempt to explain the underlying effective connectivity between the temporal and frontal cortices during the MMR, extending over a time window of 0-375 ms relative to the onset of the third tone in the deviant and control trials. In comparing the models, we initially used the BMS with an FFX approach to identify the model that best explained the MMR (Fig. 4A). The model consisting of the bilateral A1, bilateral STG, and right IFG, with intrinsic connections in the bilateral A1 (model 12), showed the highest log-evidence (F), representing very strong evidence relative to other models ($\Delta F > 5$, which is equivalent to a Bayes factor of $> 150$, $\Delta F$ indicates the difference in the log evidence between the winning and the second place model). The posterior probability for model 12 exceeded 0.99, demonstrating the high probability of this model, given the evidence, within the proposed model space. We repeated the analysis using an RFX approach to rule out the possibility of individual participant bias (due to the presence of participant outliers). This process again showed model 12 to be the winning model, with the highest model exceedance probability, $p = 0.52$ (Fig. 4B). This result shows the general agreement between the FFX and RFX approaches. As already mentioned, model 12 included the bilateral A1, bilateral STG, and right IFG, with intrinsic connections in the bilateral A1. The connections between the bilateral A1 and bilateral STG and those between the right STG and right IFG were bidirectional and included both forward and backward connections.

Finally, we used *post hoc* family-level inference to evaluate the importance of the backward or top-down connections between the IFG and STG sources to explain the MMR to rhythm deviation. The RFX results showed the FB model family (forward-backward), with a family exceedance probability of $p = 0.95$, to be the winning model family (Fig. 4C). This suggests that the rhythm deviant MMR in the current study requires the backward connections between the IFG and STG.

**Discussion**

Our results show that early in the course of development, soon after the onset of the establishment of thalamocortical circuits for auditory perception, the premature brain detects violations from a rhythmic structure. Our results further show that the processing of rhythm deviation is not limited to the primary auditory areas but encompasses a hierarchy of temporo-frontal cortical structures in a bottom-up and top-down stream, as in adults.

In very preterm infants younger than 26 wGA, thalamocortical afferents accumulate in the superficial subplate. Then, between 26 and 28 wGA, thalamocortical afferents invade the cortical plate of corresponding target areas, within which the first synapses appear. Between 28 and 30 wGA, thalamocortical axons establish synapses with cortical plate layer IV neurons and become functionally sensory-driven (4, 59). Although we cannot rule out auditory learning in our preterm population in the womb from the point when auditory processing becomes functional (2, 3), the presence of the capacity in premature neural networks to detect violations from a rhyhtmic structure suggests the possible genetic endowment of human premature networks with the capacity to process the rhythmic aspects of the auditory world. This also suggests that certain capacities observed in infants during the first year of life in terms of processing the basic temporal characteristics of auditory streams, including entrainment to rhythm (32) and the detection of deviations from auditory temporal regularities (10, 11, 33), may have genetic fingerprints. This is, however, not in contrast to the learning and development of environment/culture related sensitivity to more complex rhythmic structures (60-62) nor the enhanced neural processing of rhythm as a result of early training (32, 33). We created the stimuli based on a simple rhythmic structure and therefore the results presented are related to the capacity for processing the basic rhythmic characteristics of sound. This is consistent with evidence showing very early neural capacities for the perception of other basic physical characteristics of sound in preterm neonates (7, 8) and the fetus (63, 64). The detection of rhythm violation requires neural coding of the relative temporal pattern in the auditory sequence and predictive information processing concerning the timing of future events. Neuronal modeling has shown that MMN in adults can be explained by creation of the prediction error signal in layer IV as the difference between thalamic input and predictive signals arising from the inhibitory interneurons in the supragragranular layer and the synaptic weights being adjusted by NMDA-dependent plasticity (65). The mismatch response observed in this study suggests that a similar mechanism may already be present, at least to a certain extent, at the onset of the establishment of the thalamocortical and cortico-cortical circuits.

fNIRS (7) showed that the mismatch response in 30 wGA preterm neonates is not limited to the auditory cortex but also involves different areas of the perisylvian cortex, including the IFG. According to predictive coding models, the incongruence of the deviant with the higher-order cortical representation of the auditory stream creates an error signal, the processing of which involves local adaptation within the primary auditory cortices, as well as plasticity in inter-regional connections amongst multiple hierarchical levels (13, 15, 66). In addition, neural processing of the error is not limited to forward propagation in the cortical hierarchy, but is also shaped by a reciprocal cascade of cortical functions, in which top-down predictions serve to compare the sensed to the predicted bottom-up auditory input (66); backward connections deliver predictions to lower levels (67), whereas forward connections transfer prediction errors to upper levels (15, 67). The current study suggests that the dialog in the ventral pathway in the temporo-frontal network during the timing of the mismatch response encompasses a bottom-up and top-down stream, even in preterm neonates, as the dialog involves both forward and backward connectivity between the temporal and frontal cortices, as observed in adults (18, 42, 48, 49). Feedforward connectivity is established prenatally in primates, whereas feedback connectivity has been demonstrated to go through protracted remodeling to resemble adult-like connectivity (68). The relatively prolonged maturation of feedback connectivity does not impede their functionality and participation in the processing of predictions and prediction errors. Future studies are required to address the role of structural maturation on the long-distance connections of the DCM model that best explain the mismatch response.

The neurodevelopment literature suggests a different early developmental time-course for the left and right hemispheres. Sulci generally develop earlier in the right than left hemisphere (69). Within the perisylvian cortical areas, more advanced maturational indices have been observed in the right hemisphere for inferior frontal and superior temporal gyri and in the left hemisphere for angular and middle temporal gyri (70). Leroy et al. (71) demonstrated rightward STS asymmetry in the maturation indices, but a reversed pattern in Broca's area. Overall, fMRI and fNIRS studies have suggested greater activation in response to speech in the left than right temporal gyri and in the right than left frontal cortex, including the IFG, in preterm newborns, as well as during the first post-natal months (7, 72, 73). In an fMRI study, general activation with right-hemispheric dominance was observed in the primary, secondary, and higher order auditory cortices in newborns in response to music (37). Functional or structural developmental asymmetry indices in favor of the left or right hemisphere in different studies do not reject the possible involvement of the two hemispheres in different cognitive tasks, even in early developmental stages. The role of the right IFG in the processing of the mismatch response and its top-down modulatory role in preterm newborns in this study is in agreement with DCM studies on the MMN (40, 42, 49), and fMRI studies in adults (74-77), in which stronger activation was observed in the right IFG during the MMN. In contrast to our results, Basirat et al. (38) suggested stronger top-down modulation from higher-level regions for the left than right hemisphere in four-month-old infants after observing a late frontal negative slow wave over the left inferior frontal region. Aside from the age of the subjects, as well as the nature of the experimental protocol, one reason for this difference may be related to the different timing of the neural response being considered. We performed the DCM over the MMR (0-375 ms), whereas the left frontal slow wave in (38) was observed over 900-1200 ms. Although a late deflection was observed in the neural response corresponding to the deviant condition, the difference from the control condition did not reach significance, probably due to the oddball nature of the paradigm, as well as the duration of the trials. This could have resulted in masking of the late slow response in the tone response corresponding to the following trial.

The choice of the active sources in the hierarchical prediction in the adult temporo-frontal cortex comes from functional brain imaging, in which deviant auditory stimuli were shown to evoke neural responses in the bilateral auditory cortex, superior temporal gyri, and prefrontal cortex (74-78). Building on these studies, Garrido et al. (79) found clear evidence for a temporo-frontal hierarchy of prediction and transmission of the prediction error message using DCM and later replicated the proposed model in multiple studies (49, 66), modeling the right prefrontal cortical sources. Some other previous studies also modeled unilateral prefrontal cortical sources (40, 80), whereas others used bilateral sources (18, 48, 81). We quantitatively compared right and left unilateral with bilateral models and found very strong evidence in favor of right unilateral frontal cortical sources in the premature neonatal network. Whether the higher posterior probability of the model with the right unilateral frontal cortical source is related to the simple structure of the rhythmic sequence needs to be investigated in future studies, with modulation of the complexity of the rhythmic structure. In addition, whether the structure of the winning model varies during the course of development and through maturation of the frontal cortex, as well as the long distance connections, has to be determined in future studies.

DCM enabled us to test our specific hypothesis concerning the contribution of the frontal cortex and backward connections to the deviant rhythm response. DCM is based on local interactions between excitatory and inhibitory connections and long-distance excitatory connections (54). The neural network dynamics, as well as the local excitatory-inhibitory interactions, are not exactly the same between preterm neonates and adults. During the late preterm period, the laminar structure is still developing and long associative pathways also undergo intense development. This period is also marked by significant dendritic differentiation and synaptogenesis (4). In addition, generalization from animal studies suggests that the development of inhibitory GABA and its switch from depolarization to hyperpolarization is ongoing in the cortex (82-84). However, evidence suggests that the long distance connections are already in place (4, 68) and that the ventral pathway is functional (85, 86). The interneurons, which play an important role in the local circuitry, have been observed at 32 wGA (87) and postportum studies in seven-month old fetuses showed an established layer IV, with typical stellate cells, as well as both superficial and deep pyramidal cells (88). In light of the already discussed evidence, simplified local and distant interactions in DCM can be assumed for late preterm infants, however this extrapolation has to be treated with caution. In this study, the model parameters where those defined by default in DCM in adults (49), as there is no physiological data corresponding to the degree of variation in preterm infants. However, we considered the different neural structures in preterm infants by changing the forward model to a model considering the preterm head and structural layers (52, 53). It is less likely that slight variations in model parameters change the involvement of the frontal cortex or backward connections in the winning model corresponding to the MMR. However this has to be verified in future studies. Despite these possible limitations, the use of DCM and high-resolution ERPs in the current study has provided new information about the cortical mechanisms underlying rhythm processing in the premature brain.

In conclusion, this study provides evidence that soon after connection of the thalamocortical afferents to the cortical plate, the preterm neonatal network is able to detect deviations from a rhythmic structure. In addition, we show that this mechanism (or processing) cannot be explained by local adaptation at the auditory sensory cortex only, but results from more complex mechanisms involving higher level cortical frontal areas in a bottom-up and top-down stream. The role of these higher-level cortical areas may be different, or at least modulated in preterm infants at risk of developing neurodevelopmental disorders, given their role in supporting higher cognitive functions. Therefore, individual-level modeling and association of the results with other neurobiomarkers (notably those extracted from endogenous neural activity) may have predictive value in this at risk population.

**References**
1. M. CHEOUR-LUHTANEN *et al.* (1996) The ontogenetically earliest discriminative response of the human brain. (Wiley Online Library).
2. S. Ullal, C. M. Vanden Bosch der Nederlanden, P. Tichko, A. Lahav, E. E. Hannon, Linking prenatal experience to the emerging musical mind. *Frontiers in systems neuroscience* **7**, 48 (2013).
3. J. Gervain, The role of prenatal experience in language development. *Current opinion in behavioral sciences* **21**, 62-67 (2018).
4. I. Kostović, G. Sedmak, M. Judaš, Neural histology and neurogenesis of the human fetal and infant brain. *NeuroImage* **188**, 743-773 (2019).
5. J. Dubois, I. Kostovic, M. Judas (2015) Development of structural and functional connectivity.
6. E. Takahashi, R. D. Folkerth, A. M. Galaburda, P. E. Grant, Emerging cerebral connectivity in the human fetal brain: an MR tractography study. *Cerebral cortex* **22**, 455-464 (2012).
7. M. Mahmoudzadeh *et al.*, Syllabic discrimination in premature human infants prior to complete formation of cortical layers. *Proceedings of the National Academy of Sciences* **110**, 4846-4851 (2013).
8. M. Mahmoudzadeh, F. Wallois, G. Kongolo, S. Goudjil, G. Dehaene-Lambertz, Functional maps at the onset of auditory inputs in very early preterm human neonates. *Cerebral Cortex* **27**, 2500-2512 (2017).
9. L. J. Trainor, K. A. Corrigall, "Music acquisition and effects of musical experience" in Music perception. (Springer, 2010), pp. 89-127.

10. G. P. Háden, H. Honing, M. Török, I. Winkler, Detecting the temporal structure of sound sequences in newborn infants. *International Journal of Psychophysiology* **96**, 23-28 (2015).
11. I. Winkler, G. P. Háden, O. Ladinig, I. Sziller, H. Honing, Newborn infants detect the beat in music. *Proceedings of the National Academy of Sciences* **106**, 2468-2471 (2009).
12. K. Friston, Beyond phrenology: what can neuroimaging tell us about distributed circuitry? *Annual review of neuroscience* **25**, 221-250 (2002).
13. K. Friston, A theory of cortical responses. *Philosophical transactions of the Royal Society B: Biological sciences* **360**, 815-836 (2005).
14. M. Heilbron, M. Chait, Great expectations: is there evidence for predictive coding in auditory cortex? *Neuroscience* **389**, 54-73 (2018).
15. S. Koelsch, P. Vuust, K. Friston, Predictive processes and the peculiar case of music. *Trends in Cognitive Sciences* **23**, 63-77 (2019).
16. P. Vuust, M. J. Dietz, M. Witek, M. L. Kringelbach, Now you hear it: A predictive coding model for understanding rhythmic incongruity. *Annals of the New York Academy of Sciences* **1423**, 19-29 (2018).
17. R. Auksztulewicz *et al.*, Not all predictions are equal:"what" and "when" predictions modulate activity in auditory cortex through different mechanisms. *Journal of Neuroscience* **38**, 8680-8693 (2018).
18. H. N. Phillips, A. Blenkmann, L. E. Hughes, T. A. Bekinschtein, J. B. Rowe, Hierarchical organization of frontotemporal networks for the prediction of stimuli across multiple dimensions. *Journal of Neuroscience* **35**, 9255-9264 (2015).
19. B. Morillon, S. Baillet, Motor origin of temporal predictions in auditory attention. *Proceedings of the National Academy of Sciences* **114**, E8913-E8921 (2017).
20. E. Geiser, E. Ziegler, L. Jancke, M. Meyer, Early electrophysiological correlates of meter and rhythm processing in music perception. *cortex* **45**, 93-102 (2009).
21. P. Vuust, L. Liikala, R. Näätänen, P. Brattico, E. Brattico, Comprehensive auditory discrimination profiles recorded with a fast parametric musical multi-feature mismatch negativity paradigm. *Clinical Neurophysiology* **127**, 2065-2077 (2016).
22. E. S. Lelo-de-Larrea-Mancera, Y. Rodríguez-Agudelo, R. Solís-Vivanco, Musical rhythm and pitch: A differential effect on auditory dynamics as revealed by the N1/MMN/P3a complex. *Neuropsychologia* **100**, 44-50 (2017).
23. C. Lappe, M. Lappe, C. Pantev, Differential processing of melodic, rhythmic and simple tone deviations in musicians-an MEG study. *NeuroImage* **124**, 898-905 (2016).
24. H. Honing, O. Ladinig, G. P. Háden, I. Winkler, Is beat induction innate or learned? Probing emergent meter perception in adults and newborns using event-related brain potentials. *Annals of the New York Academy of Sciences* **1169**, 93-96 (2009).
25. J. A. Grahn, Neural mechanisms of rhythm perception: current findings and future perspectives. *Topics in cognitive science* **4**, 585-606 (2012).
26. F. L. Bouwer, H. Honing, Temporal attending and prediction influence the perception of metrical rhythm: evidence from reaction times and ERPs. *Frontiers in psychology* **6**, 1094 (2015).
27. C. Lappe, O. Steinsträter, C. Pantev, Rhythmic and melodic deviations in musical sequences recruit different cortical areas for mismatch detection. *Frontiers in Human Neuroscience* **7**, 260 (2013).
28. F. L. Bouwer, T. L. Van Zuijen, H. Honing, Beat processing is pre-attentive for metrically simple rhythms with clear accents: an ERP study. *PloS one* **9**, e97467 (2014).
29. T. C. Zhao, H. G. Lam, H. Sohi, P. K. Kuhl, Neural processing of musical meter in musicians and non-musicians. *Neuropsychologia* **106**, 289-297 (2017).

30. M. Edalati, M. Mahmoudzadeh, J. Safaie, F. Wallois, S. Moghimi, Great expectations in music: violation of rhythmic expectancies elicits late frontal gamma activity nested in theta oscillations. *arXiv preprint arXiv:2011.12676* (2020).
31. J. Moser *et al.*, Magnetoencephalographic signatures of hierarchical rule learning in newborns. *Developmental cognitive neuroscience* **46**, 100871 (2020).
32. L. K. Cirelli, C. Spinelli, S. Nozaradan, L. J. Trainor, Measuring neural entrainment to beat and meter in infants: effects of music background. *Frontiers in neuroscience* **10**, 229 (2016).
33. T. C. Zhao, P. K. Kuhl, Musical intervention enhances infants' neural processing of temporal structure in music and speech. *Proceedings of the National Academy of Sciences* **113**, 5212-5217 (2016).
34. M. H. Thaut, P. D. Trimarchi, L. M. Parsons, Human brain basis of musical rhythm perception: common and distinct neural substrates for meter, tempo, and pattern. *Brain sciences* **4**, 428-452 (2014).
35. M. Lumaca, M. J. Dietz, N. C. Hansen, D. R. Quiroga-Martinez, P. Vuust, Perceptual learning of tone patterns changes the effective connectivity between Heschl's gyrus and planum temporale. *Human Brain Mapping* **42**, 941-952 (2021).
36. A. Adam-Darque *et al.*, Neural correlates of voice perception in newborns and the influence of preterm birth. *Cerebral Cortex* **30**, 5717-5730 (2020).
37. D. Perani *et al.*, Functional specializations for music processing in the human newborn brain. *Proceedings of the National Academy of Sciences* **107**, 4758-4763 (2010).
38. A. Basirat, S. Dehaene, G. Dehaene-Lambertz, A hierarchy of cortical responses to sequence violations in three-month-old infants. *Cognition* **132**, 137-150 (2014).
39. M. I. Garrido, J. M. Kilner, S. J. Kiebel, K. J. Friston, Evoked brain responses are generated by feedback loops. *Proceedings of the National Academy of Sciences* **104**, 20961-20966 (2007).
40. M. Boly *et al.*, Preserved feedforward but impaired top-down processes in the vegetative state. *Science* **332**, 858-862 (2011).
41. S. J. Kiebel, M. I. Garrido, R. J. Moran, K. J. Friston, Dynamic causal modelling for EEG and MEG. *Cognitive neurodynamics* **2**, 121 (2008).
42. M. I. Garrido, J. M. Kilner, S. J. Kiebel, K. E. Stephan, K. J. Friston, Dynamic causal modelling of evoked potentials: a reproducibility study. *Neuroimage* **36**, 571-580 (2007).
43. M. Kleiner, D. Brainard, D. Pelli, What's new in Psychtoolbox-3? (2007).
44. R. Oostenveld, E. Maris, J.-M. Schoffelen, FieldTrip: Open Source Software for Advanced Analysis of MEG, EEG, and Invasive Electrophysiological Data. *Computational Intelligence and Neuroscience* **2011**, 41-49 (2011).
45. A. Delorme, S. Makeig, EEGLAB: an open source toolbox for analysis of single-trial EEG dynamics including independent component analysis. *Journal of neuroscience methods* **134**, 9-21 (2004).
46. C. He, L. Hotson, L. J. Trainor, Mismatch responses to pitch changes in early infancy. *Journal of cognitive neuroscience* **19**, 878-892 (2007).
47. E. Maris, R. Oostenveld, Nonparametric statistical testing of EEG-and MEG-data. *Journal of neuroscience methods* **164**, 177-190 (2007).
48. S. Chennu *et al.*, Silent expectations: dynamic causal modeling of cortical prediction and attention to sounds that weren't. *Journal of Neuroscience* **36**, 8305-8316 (2016).
49. M. I. Garrido, J. M. Kilner, S. J. Kiebel, K. J. Friston, Dynamic causal modeling of the response to frequency deviants. *Journal of Neurophysiology* **101**, 2620-2631 (2009).
50. G. Cooray, M. Garrido, T. Brismar, L. Hyllienmark, The maturation of mismatch negativity networks in normal adolescence. *Clinical Neurophysiology* **127**, 520-529 (2016).
51. K. M. Larsen *et al.*, Altered auditory processing and effective connectivity in 22q11. 2 deletion syndrome. *Schizophrenia research* **197**, 328-336 (2018).

52. S. Ghadimi, H. A. Moghaddam, R. Grebe, F. Wallois, Skull segmentation and reconstruction from newborn CT images using coupled level sets. *IEEE journal of biomedical and health informatics* **20**, 563-573 (2015).
53. H. Azizollahi, A. Aarabi, F. Wallois, Effect of structural complexities in head modeling on the accuracy of EEG source localization in neonates. *Journal of Neural Engineering* **17**, 056004 (2020).
54. O. David *et al.*, Dynamic causal modeling of evoked responses in EEG and MEG. *NeuroImage* **30**, 1255-1272 (2006).
55. S. J. Kiebel, O. David, K. J. Friston, Dynamic causal modelling of evoked responses in EEG/MEG with lead field parameterization. *NeuroImage* **30**, 1273-1284 (2006).
56. W. D. Penny, K. E. Stephan, A. Mechelli, K. J. Friston, Comparing dynamic causal models. *Neuroimage* **22**, 1157-1172 (2004).
57. K. E. Stephan *et al.*, Ten simple rules for dynamic causal modeling. *Neuroimage* **49**, 3099-3109 (2010).
58. R. E. Kass, A. E. Raftery, Bayes factors. *Journal of the american statistical association* **90**, 773-795 (1995).
59. I. Kostović, M. Judaš, The development of the subplate and thalamocortical connections in the human foetal brain. *Acta paediatrica* **99**, 1119-1127 (2010).
60. J. E. Nave-Blodgett, J. S. Snyder, E. E. Hannon, Hierarchical beat perception develops throughout childhood and adolescence and is enhanced in those with musical training. *Journal of Experimental Psychology: General* **150**, 314 (2021).
61. G. Soley, E. E. Hannon, Infants prefer the musical meter of their own culture: a cross-cultural comparison. *Developmental psychology* **46**, 286 (2010).
62. E. E. Hannon, S. E. Trehub, Tuning in to musical rhythms: Infants learn more readily than adults. *Proceedings of the National Academy of Sciences* **102**, 12639-12643 (2005).
63. R. Draganova *et al.*, Sound frequency change detection in fetuses and newborns, a magnetoencephalographic study. *Neuroimage* **28**, 354-361 (2005).
64. J. Muenssinger *et al.*, Auditory habituation in the fetus and neonate: an fMEG study. *Developmental science* **16**, 287-295 (2013).
65. C. Wacongne, J.-P. Changeux, S. Dehaene, A neuronal model of predictive coding accounting for the mismatch negativity. *Journal of Neuroscience* **32**, 3665-3678 (2012).
66. M. I. Garrido, J. M. Kilner, K. E. Stephan, K. J. Friston, The mismatch negativity: a review of underlying mechanisms. *Clinical neurophysiology* **120**, 453-463 (2009).
67. H.-J. Park, K. Friston, Structural and functional brain networks: from connections to cognition. *Science* **342** (2013).
68. H. Kennedy, R. Douglas, K. Knoblauch, C. Dehay (2007) Self-organization and pattern formation in primate cortical networks. in *Novartis Foundation Symposium* (Wiley Online Library), p 178.
69. J. Dubois *et al.*, Primary cortical folding in the human newborn: an early marker of later functional development. *Brain* **131**, 2028-2041 (2008).
70. P. Adibpour, J. Lebenberg, C. Kabdebon, G. Dehaene-Lambertz, J. Dubois, Anatomo-functional correlates of auditory development in infancy. *Developmental cognitive neuroscience* **42**, 100752 (2020).
71. F. Leroy *et al.*, Early maturation of the linguistic dorsal pathway in human infants. *Journal of Neuroscience* **31**, 1500-1506 (2011).
72. G. Dehaene-Lambertz *et al.*, Functional organization of perisylvian activation during presentation of sentences in preverbal infants. *Proceedings of the National Academy of Sciences* **103**, 14240-14245 (2006).
73. S. Shultz, A. Vouloumanos, R. H. Bennett, K. Pelphrey, Neural specialization for speech in the first months of life. *Developmental Science* **17**, 766-774 (2014).

74. S. Molholm, A. Martinez, W. Ritter, D. C. Javitt, J. J. Foxe, The neural circuitry of pre-attentive auditory change-detection: an fMRI study of pitch and duration mismatch negativity generators. *Cerebral Cortex* **15**, 545-551 (2005).
75. B. Opitz, T. Rinne, A. Mecklinger, D. Y. Von Cramon, E. Schröger, Differential contribution of frontal and temporal cortices to auditory change detection: fMRI and ERP results. *Neuroimage* **15**, 167-174 (2002).
76. C. F. Doeller *et al.*, Prefrontal cortex involvement in preattentive auditory deviance detection:: neuroimaging and electrophysiological evidence. *Neuroimage* **20**, 1270-1282 (2003).
77. T. Rinne, A. Degerman, K. Alho, Superior temporal and inferior frontal cortices are activated by infrequent sound duration decrements: an fMRI study. *Neuroimage* **26**, 66-72 (2005).
78. S. Chennu *et al.*, Expectation and attention in hierarchical auditory prediction. *Journal of Neuroscience* **33**, 11194-11205 (2013).
79. M. I. Garrido *et al.*, The functional anatomy of the MMN: a DCM study of the roving paradigm. *Neuroimage* **42**, 936-944 (2008).
80. A. Schmidt *et al.*, Modeling ketamine effects on synaptic plasticity during the mismatch negativity. *Cerebral Cortex* **23**, 2394-2406 (2013).
81. L. E. Hughes, B. C. Ghosh, J. B. Rowe, Reorganisation of brain networks in frontotemporal dementia and progressive supranuclear palsy. *NeuroImage: Clinical* **2**, 459-468 (2013).
82. Y. Murata, M. T. Colonnese, Thalamic inhibitory circuits and network activity development. *Brain research* **1706**, 13-23 (2019).
83. M. T. Colonnese, M. A. Phillips, Thalamocortical function in developing sensory circuits. *Current opinion in neurobiology* **52**, 72-79 (2018).
84. Y. Ben-Ari, Excitatory actions of gaba during development: the nature of the nurture. *Nature Reviews Neuroscience* **3**, 728-739 (2002).
85. J. Brauer, A. Anwander, D. Perani, A. D. Friederici, Dorsal and ventral pathways in language development. *Brain and language* **127**, 289-295 (2013).
86. D. Perani *et al.*, Neural language networks at birth. *Proceedings of the National Academy of Sciences* **108**, 16056-16061 (2011).
87. L. Mrzljak, H. B. Uylings, I. Kostovic, C. G. van Eden, Prenatal development of neurons in the human prefrontal cortex: I. A qualitative Golgi study. *Journal of comparative neurology* **271**, 355-386 (1988).
88. M. Marin-Padilla, Prenatal and early postnatal ontogenesis of the human motor cortex: a Golgi study. I. The sequential development of the cortical layers. *Brain research* **23**, 167-183 (1970).

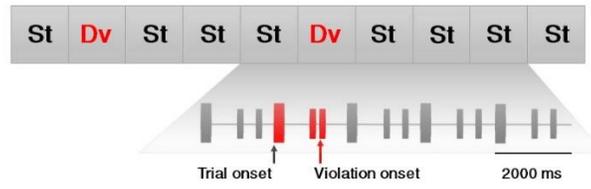
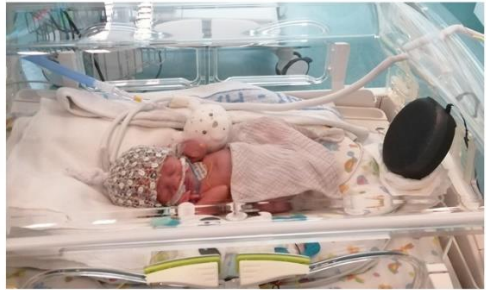

**Figure 1. (A) Scheme of the experimental protocol and stimuli. (B) Preterm neonate during the experiment with high-density EEG cap.**

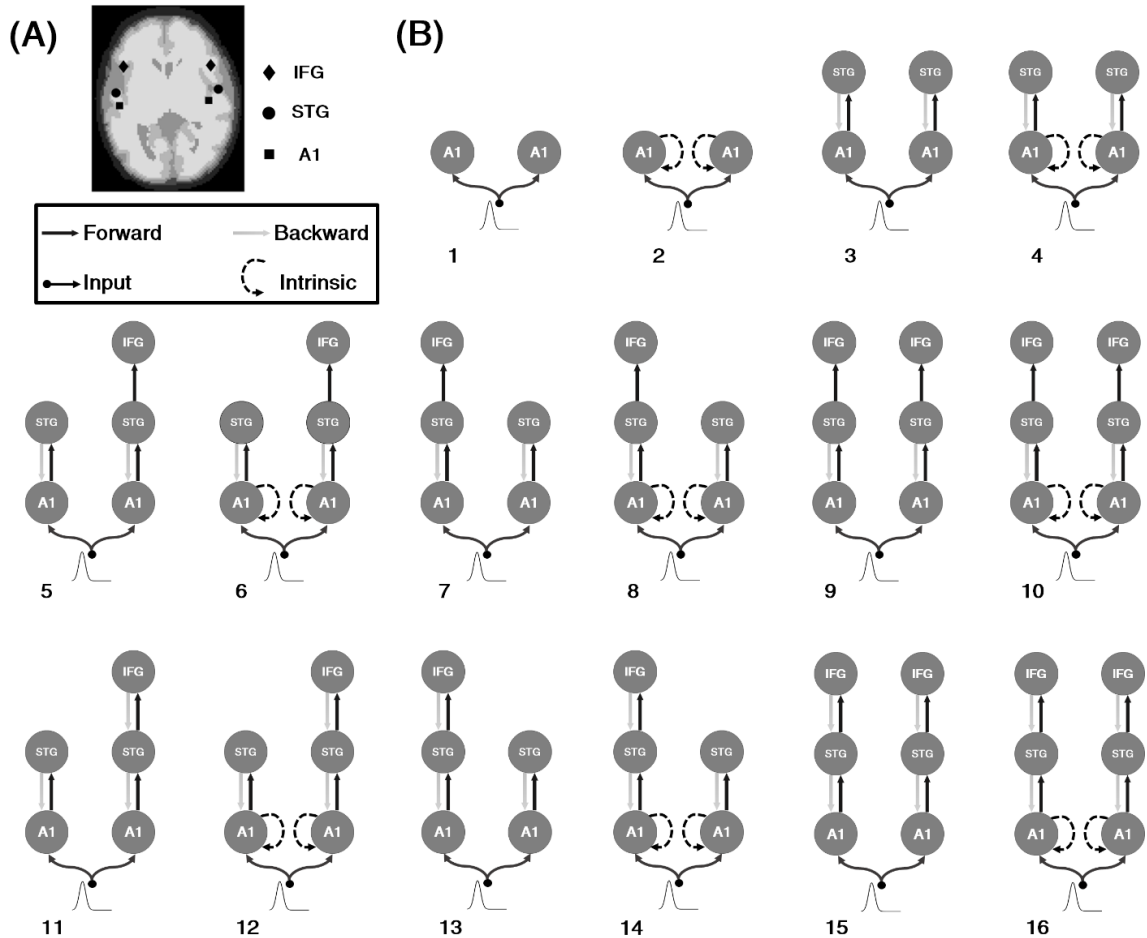

**Figure 2.** (A) Source locations included in the DCM analysis superimposed over the MRI image of a preterm neonate. (B) Dynamic causal models are designed to model the mismatch response in preterm neonates. The sources comprising the models (A1, primary auditory cortex; STG, superior temporal gyrus; IFG, inferior temporal gyrus) are connected by forward (black), backward (light grey), and intrinsic (dashed line) connections. The first four networks (NI family) ignored the role of the IFG to model the mismatch response, whereas models 5 to 10 (F family) included the IFG only, with forward connections. The last six models (FB family) considered the backward connections, in addition to the forward connections, to model the mismatch response in preterm neonates.

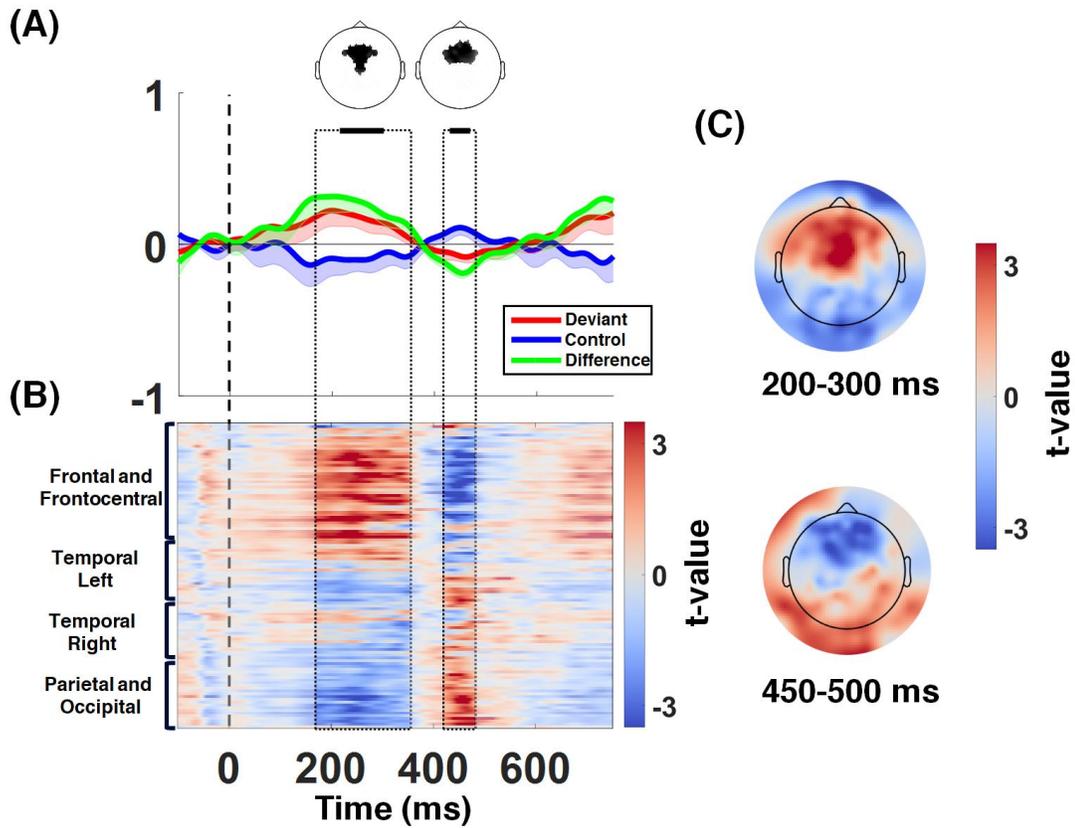

**Figure 3.** Grand-averaged event-related potentials (ERPs, averaged over all subjects) in response to the deviant and control rhthyms. The onset of the deviant was set to 0 and the last 100 ms before the onset of the deviant tone was considered to be the baseline. (A) Grand average of ERPs (-SE) for the deviant rhythm condition, the control condition, and their difference over the frontal and fronto-central electrodes. The deviant rhythm elicited an MMR, followed by a subsequent negative deflection. The black bars and head maps over the ERP figure represent the time intervals and electrodes that were significatnly different between the deviant and control conditions (p < 0.05, corrected, marked according to cluster-based permutation analysis). (B) Distribution of t-values over electrodes and time epochs. (C) The topographical distribution of t-values over significant time windows.

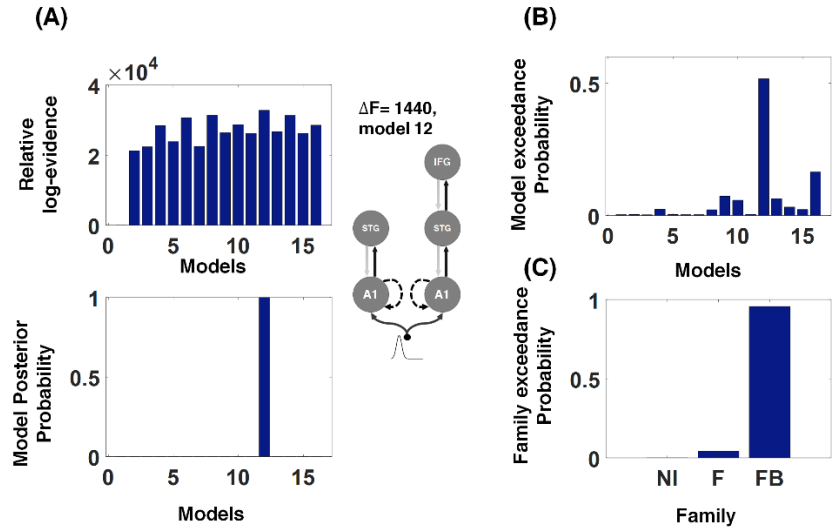

Figure 4. Bayesian model selection over the 16 tested networks. (A) Relative log-evidence and posterior probability results from FFX Bayesian model selection for each model to show their ability to model the MMR effect. The winning model (model 12), according to the results of FFX Bayesian model selection. ΔF indicates the difference between the two highest log-evidence models and a $\Delta F = 1440$, as found for model 12, is conventionally considered to be very strong evidence. (B) RFX Bayesian model selection showed model 12 to have the greatest exceedance probability and confirmed the FFX results. (C) Family-wise model selection using the RFX approach. The investigated model families: NI, no IFG models (1 - 4); F, models with only forward connections between the STG and IFG (5 - 10); FB, models with forward and backward connections between the STG and IFG (11 - 16).

Table S1. Clinical features of the tested neonates

| Neonate no. | Sex | GA at birth (wk) | GA at test (wk) | Birth weight (g) | Apgar (1 min) | Apgar (5 min) | Brain US | EEG Cap | Delivery | Presentation |
|---|---|---|---|---|---|---|---|---|---|---|
| 1 | F | 32 2/7 | 34 3/7 | 1605 | 6 | 7 | Normal | Normal | Cesarean | Cephalic |
| 2 | F | 32 3/7 | 33 6/7 | 1800 | 8 | 9 | Normal | Normal | Vaginal | Cephalic |
| 3 | F | 32 3/7 | 34 | 1690 | 10 | 10 | Normal | Normal | Vaginal | Cephalic |
| 4 | F | 31 4/7 | 34 | 1400 | 6 | 6 | Normal | Normal | Vaginal | Cephalic |
| 5 | F | 31 4/7 | 34 2/7 | 1465 | 10 | 10 | Normal | Normal | Vaginal | Cephalic |
| 6 | F | 32 1/7 | 33 6/7 | 1625 | 10 | 10 | Normal | Normal | Vaginal | Cephalic |
| 7 | F | 33 4/7 | 34 5/7 | 1840 | 10 | 10 | Normal | Normal | Vaginal | Cephalic |
| 8 | M | 31 6/7 | 33 1/7 | 1650 | 8 | 8 | Normal | Normal | Vaginal | Cephalic |
| 9 | M | 32 6/7 | 34 5/7 | 2080 | 9 | 9 | Normal | Normal | Cesarean | Cephalic |
| 10 | M | 31 | 32 5/7 | 1700 | 7 | 8 | Normal | Normal | Cesarean | Breech |
| 11 | M | 31 2/7 | 32 4/7 | 1800 | 8 | 10 | Normal | Normal | Cesarean | Breech |
| 12 | M | 31 2/7 | 32 4/7 | 1450 | 10 | 10 | Normal | Normal | Cesarean | Breech |
| 13 | M | 31 2/7 | 32 5/7 | 1460 | 6 | 7 | Normal | Normal | Cesarean | Breech |
| 14 | M | 34 | 34 6/7 | 1990 | 10 | 10 | Normal | Normal | Cesarean | Cephalic |
| 15 | F | 31 | 32 4/7 | 1660 | 8 | 8 | Normal | Normal | Cesarean | Breech |
| 16 | F | 30 | 31 4/7 | 1320 | 9 | 9 | Normal | Normal | Cesarean | Breech |
| 17 | F | 30 | 31 5/7 | 1550 | 10 | 10 | Normal | Normal | Cesarean | Breech |
| 18 | F | 29 1/7 | 30 1/7 | 1040 | 5 | 6 | Normal | Normal | Vaginal | Breech |
| 19 | M | 29 5/7 | 30 2/7 | 1100 | 6 | 7 | Normal | Normal | Cesarean | Breech |
| 20 | M | 30 2/7 | 30 5/7 | 1350 | 10 | 10 | Normal | Normal | Vaginal | Breech |

Brain US, brain ultrasonography; F, female; M, male